\documentclass[APS,PRB,twocolumn,showpacs]{revtex4}
\usepackage{amssymb}
\usepackage{txfonts}
\usepackage{graphicx}

\begin{document}

\title[]{Superconductivity  Appears in the Vicinity of an Insulating-Like Behavior in CeO$_{1-x}$F$_{x}$BiS$_{2}$  }

\author{Jie Xing, Sheng Li, Xiaxing Ding, Huang Yang  and Hai-Hu Wen}\email{hhwen@nju.edu.cn}

\affiliation{ National Laboratory of Solid State Microstructures
and Department of Physics, Center for Superconducting Physics and
Materials, Nanjing University, Nanjing 210093, China}

\begin{abstract}
Resistive and magnetization properties have been measured in
BiS$_2$-based samples CeO$_{1-x}$F$_{x}$BiS$_{2}$ with a
systematic substitution of O with F (0 $<$ x $<$ 0.6). In contrast
to the band structure calculations, it is found that the parent
phase of CeOBiS$_2$ is a bad metal, instead of an band insulator.
By doping electrons into the system, it is surprising to find that
superconductivity appears together with an insulating normal
state. This evolution is clearly different from the cuprate and
the iron pnictide systems, and is interpreted as approaching the
von Hove singularity.  Furthermore, ferromagnetism which may arise
from the Ce moments, has been observed in the low temperature
region in all samples, suggesting the co-existence of
superconductivity and ferromagnetism in the superconducting
samples.
\end{abstract}

 \maketitle
\maketitle

In the past decades, several new superconducting systems with
layered structures have been
discovered\cite{cuprates,organic,MgB2,iron}. Empirically it is
even anticipated that the exotic superconductivity may be achieved
with the layered, tetragonal or orthorhombic structures of the
compounds containing the 3d or 4d transition metals, because the
correlation effect is somehow strong in these type of samples. In
this context, the cuprates and the iron pnictides/chalcogenides
are typical examples. In both systems, the parent phase is either
a Mott insulator, like in the cuprates, or a bad metal, like in
the iron pnictides/chalcogenides. Through doping charges, the
electric conduction of the samples becomes much improved and
superconductivity sets in gradually. At the optimally doping point
where the superconducting transition temperature is the highest,
the resistivity exhibits normally as a metallic behavior, and some
times a linear temperature dependence of resistivity shows up as
an evidence of quantum criticality. Quite recently, Mizuguchi et
al. discovered the novel BiS$_2$-based superconductor
Bi$_4$O$_4$S$_3$ (named as 443 system) with T$^{onset}_c$ = 8.6
K\cite{BiOS}. This material has the BiS$_2$ layer with I4/mmm
structure.  About several days later, another BiS$_2$-based
superconductor, namely LaO$_{1-x}$F$_{x}$BiS$_{2}$ (named as 1112
system) was reported\cite{LaOFBiS}. Using transport and magnetic
measurements, we concluded the multiband and exotic
superconductivity in Bi$_4$O$_4$S$_3$\cite{lisheng}. This is
interesting and unexpected, people are curious to know what
induces the exotic superconductivity here. Using the high pressure
synthesizing method, it was found that $T_c$ can reach 10.6 K in
LaO$_{1-x}$F$_{x}$BiS$_{2}$\cite{LaOFBiS}. At the meantime other
groups repeat the discovery of superconductivity in the BiS$_2$
based systems\cite{lisheng,sunyp,awana}. By replacing the La with
Nd, superconductivity was also discovered at about T$_c^{onset}$ =
5.6 K\cite{Nd}. A scrutiny on the structures of all these samples
finds that the BiS$_2$ layers may be the common superconducting
planes in the compounds with many different blocking layers. The
first principles band structure calculation indicated that the
superconductivity was derived from the Bi 6$p_x$ and 6$p_y$
orbitals and might be related to the strong nesting effect of the
Fermi surface and quasi-one-dimensional bands\cite{LDA}. Pressure
experiment has been done on Bi$_4$O$_4$S$_3$ and
LaO$_{1-x}$F$_{x}$BiS$_{2}$\cite{pressure} samples, and the
results indicate that the Fermi surface is located in the vicinity
of some band edges leading to instability for superconductivity in
LaO$_{1-x}$F$_{x}$BiS$_{2}$. Because of this, substituting La by
other element like Ce in the BiS$_2$ based 1112 materials is
interesting to be tried since it can change the chemical pressure.
Further results from band structure calculation also show the
strong Fermi surface nesting effect\cite{wanxiangang}. Possible
pairing symmetries were also discussed based on the random phase
approximation (RPA)\cite{LDA,cal}. In this Letter, we report the
new superconductor CeO$_{1-x}$F$_{x}$BiS$_{2}$ with the typical
BiS$_2$ layer and P4/nmmz space group. It is found that the parent
phase is a bad metal, instead of a band insulator. Meanwhile the
superconductivity appears in accompanying with a normal state with
an insulating behavior, showing sharp contrast with the cuprates
and the iron pnictides.

\begin{figure}
\includegraphics[scale=0.85]{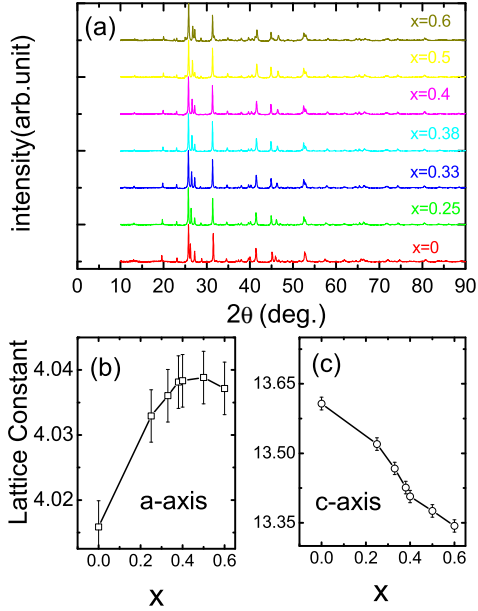}
\caption{(color online) The X-ray diffraction profile for the
powdered samples of CeO$_{1-x}$F$_{x}$BiS$_{2}$(x = 0-0.6). Except
for several minor peaks of impurities, all of peaks can be
characterized to the standard CeOBiS$_2$ with the space group
P4/nmmz.}
\end{figure}

The polycrystalline samples were grown by a conventional solid
state reaction method. First of all, we mixed Ce flakes (99.9$\%$,
Alfa Aesar), CeF$_3$ (99.9$\%$, Alfa Aesar), CeO$_2$ (99.9$\%$,
Alfa Aesar), Bi$_2$S$_3$ (99.9$\%$, Alfa Aesar) and S powder
(99.9$\%$, Alfa Aesar) by the ratio in the stoichiometry
CeO$_{1-x}$F$_{x}$BiS$_{2}$. Secondly, we pressed the mixture into
a pellet shape and sealed in an evacuated quartz tube. Then it was
heated up to 700$^\circ$C and kept for 10 h. After cooling the
compound to room temperature slowly, the product was well-mixed by
re-grinding, pressed into a pellet shape and annealed at
600$^\circ$C for 10 h. The obtained samples look black and hard.
The true composition of the samples was checked with the
Energy-dispersive X-ray spectroscopy (EDX) analysis on randomly
selected grains and found to be close to the nominal one. The
crystallinity of the sample was measured by the x-ray diffraction
(XRD) with the Brook Advanced D8 diffractometer with Cu K$\alpha$
radiation. The analysis of the XRD data was dealt with the
software Powder-X and Topas. From the PDF-2 2004, we can find that
the XRD pattern looks very similar to the  result of standard
samples of CeOBiS$_{2}$. The Rietveld fitting shows that over
90$\%$ volume of the samples are CeO$_{1-x}$F$_{x}$BiS$_{2}$ and
less than 10$\%$ are derived from the impurities which was mainly
Ce$_2$O$_2$S. As the sample is hard enough, we can cut and polish
the sample into a rectangular shape for the sequential
measurements. The resistivity and Hall effect were measured with
Quantum Design instrument PPMS-9T. The magnetization was detected
by the Quantum Design instrument SQUID-VSM with a resolution of
about 5 $\times$ 10$^{-8}$ emu. The six-lead method was applied
for the transport measurement on the longitudinal and transverse
resistivity simultaneously. The Hall effect was measured by either
sweeping magnetic field at a fixed temperature or sweeping
temperature at a fixed magnetic field. The data obtained by these
two ways seem to coincide each other.

\begin{figure}
\includegraphics[scale=0.85]{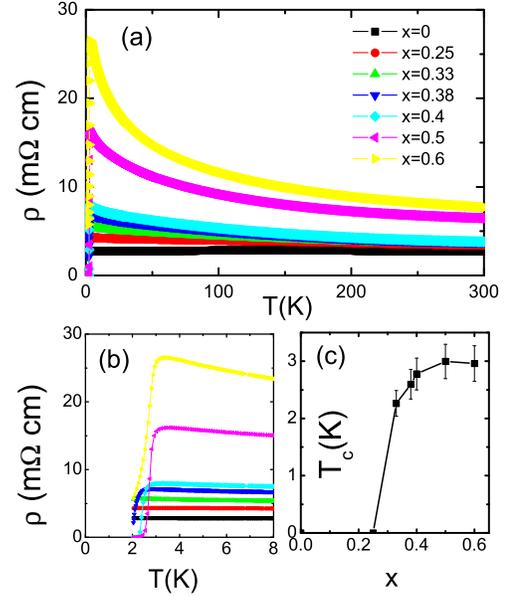}
\caption{(color online) (a) The temperature dependence of
resistivity for CeO$_{1-x}$F$_{x}$BiS$_{2}$(x = 0-0.6). It is
clear that the parent phase at x=0.0 is a bad metal, not a band
metal as expected by the LDA calculation. (b) An enlarged view of
the same data in the low temperature region. (c) The doping
dependence of the superconducting transition temperature
determined through the crossing method (see text).}
\end{figure}

\begin{figure}
\includegraphics[scale=0.75]{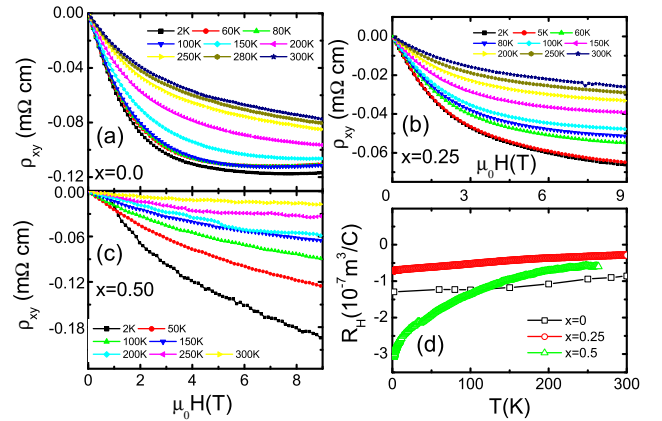}
\caption{(color online)(a) The transverse resistivity $\rho_{xy}$ versus the magnetic field $\mu_{0}H$ at 2 K, 60 K, 80 K, 100 K, 150 K, 200 K, 250 K, 280 K and 300 K for sample CeO$_{1-x}$F$_{x}$BiS$_{2}$ x = 0. (b)The transverse resistivity $\rho_{xy}$ versus the magnetic field $\mu_{0}H$ at 2 K, 5 K, 60 K, 80 K, 100 K, 150 K, 200 K, 250 K, 280 K and 300 K for sample CeO$_{1-x}$F$_{x}$BiS$_{2}$ x = 0.25. (c)The transverse resistivity $\rho_{xy}$ versus the magnetic field $\mu_{0}H$ at 2 K, 50 K, 100 K, 150 K, 200 K, 250 K, and 300 K for sample CeO$_{1-x}$F$_{x}$BiS$_{2}$ x = 0.5. (d) The Hall coefficient $R_H$ of the three samples (CeO$_{1-x}$F$_{x}$BiS$_{2}$ x=0.0, 0.25, 0.5) at 9 T from 2 K to 300 K. The dense data for x = 0.25 and 0.50 were determined by the measurement at -9 T and 9 T by sweeping temperature. The discrete data point for x=0 were determined from the data measured by sweeping the magnetic field at a fixed temperature.}
\end{figure}

Fig.~1 shows the X-ray diffraction data for the powdered samples
of CeO$_{1-x}$F$_{x}$BiS$_{2}$(x = 0-0.6). The space group of the
standard CeOBiS$_2$ is P4/nmmz with a layered structure. The XRD
pattern looks very similar to the standard CeOBiS$_2$ with a few
minor peaks of the impurity phase. The Rietveld fitting result
also reveals that the a-axis of CeO$_{1-x}$F$_{x}$BiS$_{2}$ is
4.016 $\AA$ at x = 0, increases to 4.388 $\AA$ until x = 0.5, then
it decreases to 4.037 $\AA$ at x = 0.6. The c-axis lattice
constant decreases from 13.607 $\AA$ to 13.343 $\AA$ continuously
as x is increased from x = 0 to x = 0.6. This result indicates
that the layer structure expands in the in-plane direction as more
F is doped into the system, reaches a maximum at x = 0.5, then
starts to shrink at x = 0.6. The smooth decrease of the c-axis
lattice parameter suggests that F has been successfully
substituted to the O site as the ionic radius of F is smaller than
that of O. The results seem to be similar to that of
NdO$_{1-x}$F$_{x}$BiS$_{2}$\cite{Nd}.

In Fig.~2(a) we present the temperature dependence of resistivity
for different doped sample of CeO$_{1-x}$F$_{x}$BiS$_{2}$ with x =
0-0.6.  It is clear that the parent phase CeOBiS$_2$ is not an
insulator nor a superconductor. The temperature dependence of
resistivity of parent phase presents a non-monotonic change from 2
K to 300 K. From the LDA calculation\cite{LDA}, parent phase of
this kind material should be a band insulator, being different
from the experiment result. This could be induced by two reasons:
(1) There is a self-doping in the parent phase, so that it
exhibits a metallic behavior instead of a band insulator. This is
not supported by the Hall effect measurement shown below. The Hall
data indicate that the parent phase is dominated by the
electron-charge carriers. By further doping F to O sites, one
induces more electrons into the system, therefore a better
metallic behavior should be anticipated. But actually the system
becomes more insulating-like with further doping. (2) The metallic
behavior of the parent phase may be induced by the strong
spin-orbital coupling, which shifts the bottom of the p$_x$ and
p$_y$ band below the Fermi energy. The same data with an enlarged
scale is shown in Fig.2(b), it is easy to find out that
superconductivity appears at about x = 0.33 and the transition
become sharpest at x = 0.5 with the highest transition
temperature. The onset of superconducting transition in  sample
with x = 0.5 is about 3.0 K determined by the so-called crossing
method, that is the crossing point between a normal state straight
line and an extrapolation line of the steep transition part. From
Fig.~2(b), we can also realize that the resistivity of these
materials increases with the doping of the F concentrations. It is
very strange to see that superconductivity and and an
insulating-like or semiconducting normal state appears together.
In Fig.2(c), the doping dependence of the superconducting
transition temperature is shown. It is clear that a half dome-like
superconducting area is observed here. Actually, in most 1112
samples reported so far, the superconductivity emerges on a
background of an insulating-like or a semiconducting
behavior\cite{LaOFBiS,pressure}.

In order to reveal the strange normal state behavior, we measured
the Hall effect of three samples with x = 0.0, 0.25 and 0.5.
Fig.~3(a)-(c) show the magnetic field dependence of the transverse
resistivity $\rho_{xy}$ at different temperatures of the three
samples. Fig.~3(d) shows the temperature dependence of $R_H$ of
the three samples determined at the magnetic field of 9 T.
Normally for a single band metal or a semiconductor, the Hall
coefficient $R_H$ can be measured by $R_H = d\rho_{xy}/dH = 1/n e$
with $n$ the charge carrier density when the $\rho_{xy}$ exhibits
a linear behavior with the magnetic field. While as we saw in the
previous study in the system of Bi$_4$S$_4$O$_3$, the transverse
resistivity is extremely nonlinear in magnetic field, yielding a
difficulty in determining $R_H$ in the usual way. We thus
determine $R_H$ here directly by $R_H = \rho_{xy}/H$ at 9 T. All
of these results show that the $\rho_{xy}$ of the three samples is
negative from 2 K to 300 K at 9 T, indicating the electron like
charge carriers as the dominating one. From Fig.~3(a) and (b), one
can see that the magnetic field dependence of $\rho_{xy}$ are more
curved at low doping levels. It illustrates that there may be a
very strong multi-band effect or the shallow band edge effect at
these phases. With more doping, the non-linear curvature seems a
bit weakened. In Fig.~3(d), one can clearly see that the Hall
coefficient $R_H$ of the low doped samples (x = 0.0 and 0.25) has
a weak temperature dependence. Qualitatively it is further
suggestive that more electrons are doped to the system since the
charge carrier density determined from $n=1/R_H e$ is higher in
the sample of x = 0.25 than that of x = 0.0. This may suggest that
these samples are more or less dominated by a single band at a low
doping, while with a shallow band edge, so that $\rho_{xy}$
exhibits a non-linear field dependence.  We can use the single
band assumption to estimate the charge carrier density, which is
about 10$^{19}$/cm$^3$, supporting the picture of a shallow band
edge or a small Fermi pocket. As for the sample with x = 0.5, the
Hall coefficient shows a very strong temperature dependence,
indicating that multi-scattering channels are involved.
Interestingly, the superconductivity occurs at the same time. This
suggests that the later joined scattering is very important for
superconductivity. One picture derived from the data would be that
the system is more close to the Van Hove singularity point as the
doping is close to 0.5. The LDA calculation\cite{LDA} does
indicate that the Fermi surface segments will emerge at the middle
point between $\Gamma$ (A) and the M (Z) point leading to a high
density of states peak (the von Hove Singularity effect). In this
case, a topological change of the Fermi surface is expected. It
may be this better-achieved nesting effect of the Fermi surface in
the higher doped samples that leads to a charge-density-wave (CDW)
instability, which makes the enhanced insulating background. At
the meantime, as a multi-band system often does, part of the
electrons would like to pair and condense in order to lower the
energy. The pressure study for BiS$_2$ superconductors also
elucidates that the sample with x = 0.5 is located in the vicinity
of some instability between the semiconducting and the metallic
behavior\cite{pressure}.

\begin{figure}
\includegraphics[scale=0.75]{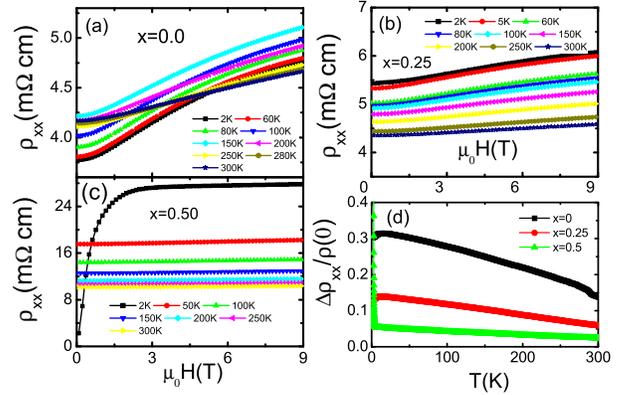}
\caption{(color online) (a) Field dependence of longitudinal resistivity $\rho_{xx}$ at 2 K, 60 K, 80 K, 100 K, 150 K, 200 K, 250 K, 280 K and 300 K for sample CeO$_{1-x}$F$_{x}$BiS$_{2}$ with x = 0.0 (b) Field dependence of longitudinal resistivity $\rho_{xx}$ at 2 K, 5 K, 60 K, 80 K, 100 K, 150 K, 200 K, 250 K, and 300 K for sample CeO$_{1-x}$F$_{x}$BiS$_{2}$ with x = 0.25. (c) Field dependence of longitudinal resistivity $\rho_{xx}$ at 2 K, 50 K, 100 K, 150 K, 200 K, 250 K, and 300 K for sample CeO$_{1-x}$F$_{x}$BiS$_{2}$ x = 0.5. (d) Temperature dependence of the magnetoresistance $\Delta\rho_{xy}$/$\rho_0$ at 9 T  for three samples of CeO$_{1-x}$F$_{x}$BiS$_{2}$ with x = 0.0, 0.25, 0.5.}
\end{figure}

\begin{figure}
\center
\includegraphics[scale=0.85]{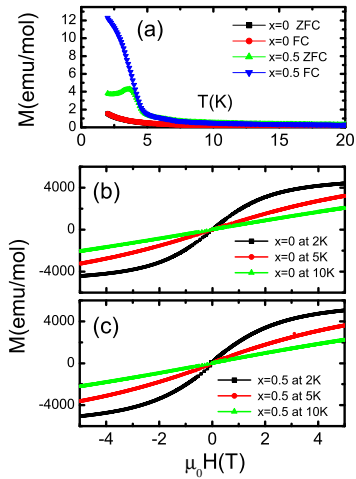}
\caption{(color online) (a) Temperature dependence of the DC
magnetization of the two samples CeO$_{1-x}$F$_{x}$BiS$_{2}$ with x = 0.0 and 0.5. (b)  Isothermal MHLs at 2 K, 5 K, 10 K of the sample CeO$_{1-x}$F$_{x}$BiS$_{2}$ x = 0.0, showing a ferromagnetic transition belwow about 6 K. (c) The MHLs measured at 2 K, 5 K and 10 K for the sample x=0.5. }
\end{figure}

Fig.~4 shows the magnetoresistance for the three typical samples
of x = 0.0, 0.25 and 0.5. From Fig.~4(a)-(c), it is easy to find
out that by increasing the electron doping, the longitudinal
resistivity $\rho_{xx}$ shows better linear character from x = 0
to x = 0.5. For the undoped sample, the $\rho_{xx}$ increases
20$\%$-30$\%$ at a magnetic field of 9 T. This is in contrast to
the sample with x = 0.5, $\rho_{xx}$ has a 5$\%$ increase at 9 T.
As the magnetoresistance of single band metal is proportional to
H$^2$ in the low field region, the magnetoresistance of
CeO$_{1-x}$F$_{x}$BiS$_{2}$ is more likely to be related to the
multiband effect at a high doping. The result is similar to the
measurements of  Bi$_4$O$_4$S$_3$\cite{lisheng}. Fig.~4(d) shows
the temperature dependence of  magnetoresistance of these three
samples at 9 T. The trend shows that with the electron doping, the
Fermi segments near ($\pm\pi$/2, $\pm\pi$/2) will show up. This
effect on one hand will lead to Van Hove singularity peak on the
DOS at the Fermi energy, on the other hand it will induce
multi-channel scattering. This conclusion is consistent with that
drawn from the Hall effect measurements. Therefore, to approach
the Von Hove singularity point and the topological change of the
Fermi surface are very important for superconductivity.

In the superconducting samples, we did not succeed in obtaining
the diamagnetism. This, at the first glance, seems in
contradiction with the conclusion of a bulk superconductor. While
a closer inspection finds that the superconducting diamagnetism is
actually prevailed over by a quite strong ferromagnetism. In
Fig.~5(a) we present the temperature dependence of magnetic
susceptibility for CeO$_{1-x}$F$_{x}$BiS$_{2}$ (x = 0.0 and 0.5)
at the field of 10 Oe. It is interesting to realize that the
parent phase has already a ferromagnetic transition at
approximately 5 K. The isothermal magntization-hysteresis-loops
(MHLs) in Fig.~5(b) give support to this conclusion as well. As
for the sample of x = 0.5, we see two steps on the
zero-field-cooled magnetization curve, one occurs at about 5 K
with a uprising of the magnetization, another one with the
relative dropping of the magnetization occurs at about 4 K. This
latter one is actually induced by the superconductivity
transition. The MHLs of the x = 0.5 sample shown in Fig.~5(c)
indicate also the dominating ferromagnetic signal. This
ferromagnetism may be induced by the local moment of Ce. For the
superconducting sample, this indicates the co-existence of
superconductivity and ferromagnetism at a low temperature. It
remains to be discovered how does the superconductivity occurring
in the BiS$_2$ layers accommodates well with the ferromagnetic
order in the CeO layer, since a bulk superconductivity requires to
establish the interlayer coupling across the ferromagnetic CeO
layers. For a singlet pairing, this seems to be challenging.

In summary, we have fabricated a new BiS$_2$-based superconducting
systems CeO$_{1-x}$F$_{x}$BiS$_{2}$ with a systematic substitution
of O with F (0.0 $<$ x $<$ 0.6). Resistivity, Hall effect,
magnetoresistance and magnetization have been conducted on them.
The parent phase is found to be a bad metal, which is not
consistent with the LDA calculations. By substituting O with more
F, superconductivity gradually appears in accompanying with an
insulating-like normal state. By analyzing the Hall effect and the
magnetoresistance and combining with the LDA calculations, we
intend to conclude that the undoped or low doped samples have a
very shallow edge with small Fermi pockets. While when it is close
to a doping level of x =0.5, the system is approaching to a Von
Hove singularity with the feature that the Fermi surface segments
near ($\pm\pi$/2, $\pm\pi$/2) will emerge. The insulating behavior
in the normal state of the superconducting sample is interpreted
as either a charge density wave instability or a gradually
enhanced correlation effect. Finally we show the coexistence of
the superconductivity with the ferromagnetic order state arising
from the local moments of Ce at low temperatures.

\begin{acknowledgments}
We appreciate the useful discussions with Liang Fu at Harvard and
Fa Wang at MIT. This work is supported by the NSF of China, the
Ministry of Science and Technology of China (973 projects:
2011CBA00102) and PAPD.

\end{acknowledgments}

\end{document}